\documentclass[aps,prl,groupedaddress,amsmath,amssymb,showpacs]{revtex4}

\usepackage{graphicx}
\usepackage{graphics}
\usepackage{dcolumn}
\usepackage{bm}

\def\lambdabarp{\lambda_p \!\!\!\!\bar{}\,\,\,\,}
\def\invomegape{\omega_{pe}^{-1}}

\def\Dbetaperp{\Delta \beta_{\perp}}

\begin{document}




\title{Inductive and Electrostatic Acceleration in Relativistic Jet-Plasma Interactions}

\author{Johnny~S.~T.~Ng and Robert~J.~Noble}

\address{
{\it Stanford Linear Accelerator Center, 
Stanford University, Stanford, CA 94309}
}


\begin{abstract}
We report on the observation of rapid particle acceleration in numerical 
simulations of relativistic jet-plasma interactions and discuss the
underlying mechanisms. The dynamics of a charge-neutral,
narrow, electron-positron jet propagating through an unmagnetized 
electron-ion plasma was investigated using a
three-dimensional, electromagnetic, particle-in-cell computer code.  
The interaction excited magnetic filamentation as well as electrostatic 
plasma instabilities.  In some cases, the longitudinal electric fields 
generated inductively and electrostatically reached the cold 
plasma wave-breaking limit, and the longitudinal momentum of about half
the positrons increased by 50\% with a maximum gain exceeding a factor of
two during the simulation period.  Particle acceleration via these mechanisms
occurred when the criteria for Weibel instability were satisfied.
  
\end{abstract}


\maketitle


Relativistic outflows are commonly associated with astrophysical sources 
such as active galactic nuclei and gamma-ray bursts.
Their interaction with ambient plasma is believed to 
give rise to particle acceleration producing
the observed radiation spectrum ranging from radio to
gamma-ray.  Acceleration occurs when the energy carried by the outflow 
is transferred to the surrounding material.  
The subject is an area of active research
and has been reviewed elsewhere \cite{Blandford94,KirkTexas04}.
In some models, the mechanism
relies on the stochastic scattering of particles among the magnetic fields
created by magnetohydrodynamic shocks \cite{KirkDuffy99}.  
Alternatively, kinetic energy of the outflow
could be converted into plasma instabilities which in turn power particle
acceleration and nonthermal radiation \cite{Schlick2002,Chen2002}.

In this Letter, we report on the observation of particle
acceleration in simulation studies of relativistic jet-plasma interactions,
and elucidate the relevant physical mechanisms.
We study the evolution of a narrow, charge-neutral, 
electron-positron jet propagating through
a stationary, unmagnetized electron-ion plasma. This geometry 
allows us to explore the physics occurring on the plasma wavelength scale 
both in the jet interior and at the jet-plasma 
boundary, in the limit of no background magnetic field. 
Our work sheds light on fundamental questions regarding
the processes by which jet-plasma interactions
cause particle acceleration.  

When charge-neutral plasmas stream through each other it is well-known that 
filamentation occurs via a process commonly referred to as the transverse 
Weibel instability \cite{weibel59,yoon87}.  
Essentially, the Lorentz force associated 
with magnetic field perturbations due to local current imbalances
causes the moving neutral plasma
to charge-separate transversely, and the resulting current filaments
strengthen the azimuthal magnetic field perturbation in a positive 
feed-back mechanism.
The result is that a neutral jet quickly breaks up into oppositely
charged filaments.  
This instability has been suggested as a mechanism
for generating strong magnetic fields in the relativistic outflows of
active galactic nuclei and
gamma-ray bursts~\cite{Medvedev99,Medvedev05}.

Particle-in-cell (PIC) simulations are well-suited to study these 
complex phenomenon~\cite{birdsall91}.
%
Recent PIC simulation
studies of filamentation in astrophysical plasmas have concentrated on
wide jets using periodic boundary conditions to study the interior 
dynamics~\cite{silva03,frederiksen04,hededal04,nishikawa05}.
Astrophysical observations have indicated instabilities might also 
be taking place at the 
boundary region of the jet, that is, the
interface of the flowing relativistic plasma and surrounding material.
It was the goal of this work to simulate the interaction of a
narrow relativistic jet where the dynamics in the interior
as well as at the jet-plasma boundary can be investigated.  
We study finite length as well as continuously flowing jets.
Our investigation of finite length jets was motivated by the observation 
of density variations along the length of some astrophysical jets, 
due to either some type of
longitudinal instability or the pulsed nature of the source.

Our PIC code is based on the TRISTAN package modified
for our particular problem~\cite{ISSS4}.  TRISTAN is a 
three-dimensional, electromagnetic, relativistic particle-in-cell code
employing a so-called local electromagnetic field solver without the
need for transform methods.
The temporal and spatial scales are normalized to $\invomegape$ and
$\lambdabarp$, respectively, of the background plasma, where
$\omega_{pe} = \sqrt{4\pi e^2 n_e / m_e}$ is the electron plasma frequency,
and
$\lambdabarp = {\rm c}/ \omega_{pe}$ is the collision-less skin depth,
$c$ being the speed of light, $e$ the electron charge, 
$n_e$ the plasma electron density, and $m_e$ the electron mass.
The electromagnetic fields are normalized to the
cold plasma-wave breaking value $E_{pw} = m_e\,c\,\omega_{pe} / e$.
Each simulation run thus represents a family of cases with arbitrary 
background plasma density because this parameter has been scaled out.
The jet's relativistic factor $\gamma$, diameter $D$, length $L$, and the jet 
to plasma
density ratio $\alpha$, as well as the time-step size and the number of 
macro-particles per cell can all be set independently.

\begin{figure}
\includegraphics[width=8.5cm]{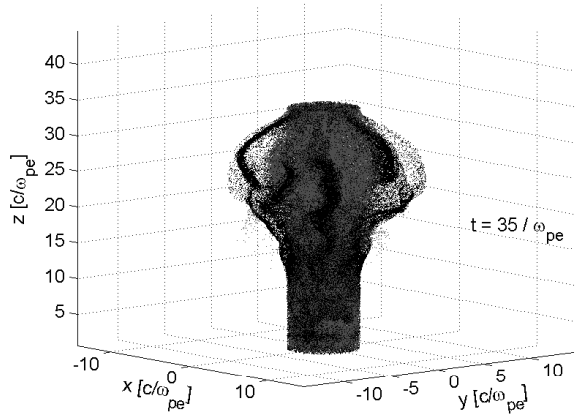}
\caption{\label{3djet} Spatial distribution of electrons (gray dots) 
and positrons (black dots) of a continuous jet.  The stationary 
background plasma is not shown.}
\end{figure}

Simulations were performed on a $150 \times 150 \times 225$ grid with a
total of about 40 million macro-particles. 
In units of $\lambdabarp$, the box size
was $30\times 30$ transversely ($x \times y$) and 45 longitudinally (in $z$).
The TRISTAN boundary conditions were set such that radiation at the box 
walls was absorbed to simulate free space with no reflections.
Particles leaving the box were lost from the 
simulation, although we recorded the energy they carried away.  
Typically the simulation box was made large enough so that few
jet particles ever leave during the entire simulation and less than $0.5\%$ of
the jet energy was carried away.
We varied our simulation parameters, and found that our results were
not sensitive to the box size, nor the number of 
macro-particles per cell once this number was in the range of 4 to 8. 

Figure~\ref{3djet} illustrates our simulation geometry and 
shows a continuous jet of uniform cross-section at
a simulation time step of $t = 35~\omega_{pe}^{-1}$.  The stationary 
electron-ion plasma fills the entire box uniformly and defines the
laboratory frame where physical parameters are given.
The jet enters at the bottom of the simulation box and travels along the
$z$ direction.  To discuss a concrete example in the following, we choose
the jet parameters to be $\gamma=10$, $\alpha=10$, 
$D = 6~\lambdabarp$, $L = 10~\lambdabarp$, and the RMS transverse
velocity spread $\Delta v_{\perp}/c = \Delta \beta_{\perp} = 10^{-4}.$
Estimates of the physical parameters for
astrophysical outflows span many orders of magnitude.  The dimensions of jets
we can simulate are much smaller than those in astrophysical environments, 
but the dynamics of interest here occur on the scale of a plasma wavelength.
The plasma physics in this example occur within a duration of 45 
$\invomegape$. 
A time-step size of $0.1~\invomegape$ was adequate 
to resolve this physics, as confirmed by runs with $0.05~\invomegape$
time-steps.  For a Courant parameter of 0.5, this corresponded 
to a mesh size of $0.2~\lambdabarp$. 
We have performed simulations with a range of parameters: $\alpha = 0.1$ to 
100,
$\Dbetaperp = 10^{-4}$ to $10^{-1}$, $D = 6$ to $60~\lambdabarp$,
and $\gamma = 10$ to 100.
We give a summary of our parameter variation studies at the end.

\begin{figure}[t]
\includegraphics[width=7.5cm]{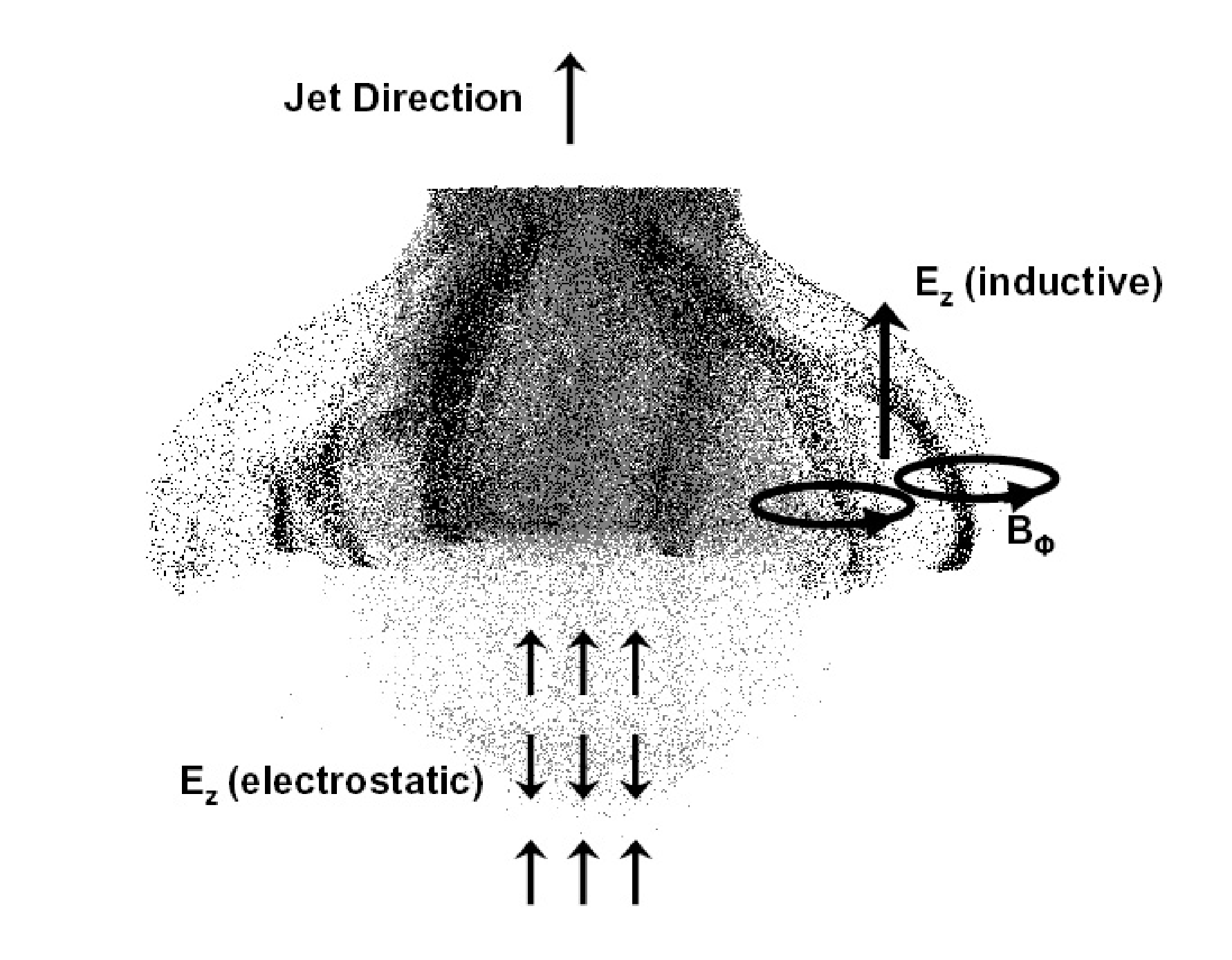}
\caption{\label{3djet_sideview_cartoon} 
Side view of a $10~\lambdabarp$ long, $6~\lambdabarp$ wide jet at 
simulation time $t = 35~\invomegape$, showing positron filaments (black dots)
expelled from the jet leaving behind the electrons (gray dots).
It also illustrates the mechanisms for generating inductive 
as well as electrostatic longitudinal electric fields.}
\end{figure}

The fundamental issue here is how jet-plasma instabilities can lead to
particle acceleration.  In our simulations, we observe the
Weibel instability as the first step in this process.
The electromagnetic energy grows
exponentially and then saturates, and eventually dissipates.
The growth rate in the laboratory frame
is approximately $\omega_{pj}/\sqrt{\gamma}$~\cite{Silva2002},
where $\omega_{pj} = \sqrt{4\pi e^2 n_j / m_e}$ is the jet's plasma
frequency, $n_j$ being the jet density.  The jet filament size is 
on the order of $c/\omega_{pj}$.
Our simulations confirm the scaling with $n_j$ and $\gamma$. 
The instability saturates
at approximately 10\% of the initial jet energy 
at $t = 20~\omega_{pe}^{-1}$, and dissipates to about
65\% of the maximum value by $t = 45~\omega_{pe}^{-1}$.

As the jet continues to plow forward,
the Lorentz force between the electron and positron filaments tend to
push them apart.  The electron filaments, however, are confined by the
electrostatic channel formed by the heavier-mass 
background plasma ions $(m_i = 192\,m_e)$.
The positron filaments are preferentially expelled from the interior of the 
jet.  The filaments near the surface escape first, followed by the interior
ones as they sequentially migrate outward.
As the positron filaments move away from the interior and from each other, the
azimuthal magnetic field $(B_\phi)$ associated with the filaments 
decreases rapidly. According to 
Faraday's law of induction, the time-variation of magnetic flux 
generates a loop-integrated electric field enclosing the flux region.
In our simulation,
we find that this large and negative $\dot{B_\phi}$ locally induces 
a large and positive longitudinal electric field $E_z$, as illustrated in
Figure~\ref{3djet_sideview_cartoon}.
Furthermore, once separated, these charged filaments also generate longitudinal
electrostatic oscillations in the background plasma.

\begin{figure}[t]
\includegraphics[width=7.5cm]{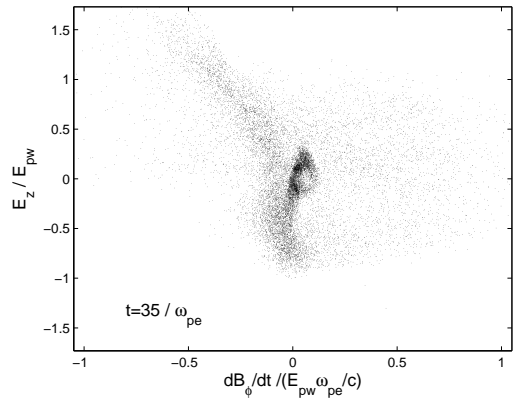}
\caption{\label{EzVsBpDot} Correlation of longitudinal electric field with
time variation of azimuthal magnetic field at the same spatial location
for a $10~\lambdabarp$ long, $6~\lambdabarp$ wide jet.  
For clarity, only a random subset 
of field points outside a radius of $2.5~\lambdabarp$ are shown.}
\end{figure}

\begin{figure}[t]
\includegraphics[width=12.5cm]{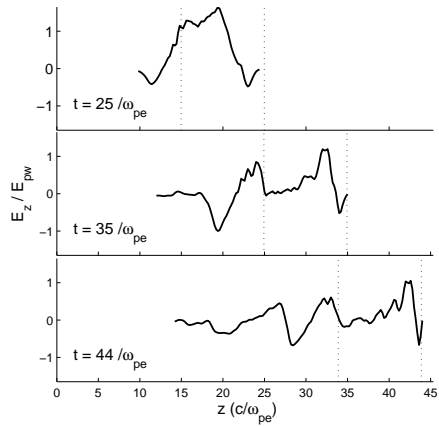}
\caption{\label{EzVsZ} Longitudinal electric field as a function of
$z$ at a given point $(x,y)=(2.25,1.25)~\lambdabarp$.  
The dotted lines bracket
the $10~\lambdabarp$ long jet.  Note the predominantly inductive $E_z$ at
$t=25~\invomegape$, and the wakefield left behind the jet at later times.}
\end{figure}

These two types of longitudinal electric fields in our simulations 
are represented by the two dominant features shown in
Figure~\ref{EzVsBpDot}.  
The inductive nature of the fields is demonstrated by
the anti-correlation
between $E_z$ and $\dot{B_\phi}$ in the upper-left quadrant of
the plot.  The electrostatic component corresponds to the cluster of
points near $\dot{B_\phi} = 0$, with both positive and negative polarities.
The waveforms of these fields are shown explicitly in Figure~\ref{EzVsZ}
at three epochs during the simulation.

The inductively generated $E_z$ field 
propagates at the same relativistic velocity as
the filaments, and persists throughout the simulation duration.
Based on experience with terrestrial accelerators, this $E_z$ can
efficiently accelerate particles as they ``surf'' on a wave of
electric fields in the direction of motion.  In this case, it preferentially
accelerates comoving positrons and decelerates electrons.

As the charge filaments separate,
we found that starting at about 25 $\invomegape$, 
a coherent train of longitudinal plasma oscillations was excited 
behind the jet analogous to the wake of a ship in water. 
They have
no associated transverse magnetic field over their interior volume, and 
only limited surface magnetic fields at their extreme edges, as expected 
for finite-size plasma waves. 
These plasma wakefields 
have a phase velocity equal to the speed of the drive jet,
and have amplitudes of order the wave-breaking 
limit for the parameters of this simulation. Plasma wakes always 
form immediately behind the trailing edge of the jet, regardless 
of its length.  They 
continue to oscillate after the drive jet passes, and hence can
accelerate relativistic particles over very long distances~\cite{Chen2002,
Esarey96}.

\begin{figure}[t]
\includegraphics[width=11cm]{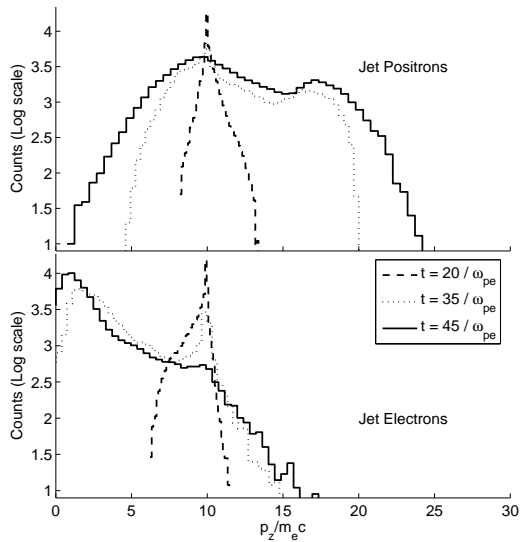}
\caption{\label{Pz_Spect} Longitudinal momentum distribution of jet positrons
and electrons at three simulation epochs for a $10~\lambdabarp$ long jet,
showing positron acceleration and electron deceleration.}
\end{figure}

The longitudinal momentum $(p_z)$ distributions of jet positrons and 
electrons are shown in Figure~\ref{Pz_Spect}.  On average the positrons
gained energy, with approximately half of the initial population
increasing at least 50\% in $p_z$.  The electrons generally lost 
energy during the process.  The positron energy change of about 
$15~m_e c^2$ during the simulation is consistent with
acceleration over a distance of 10 to 20 $\lambdabarp$
by a longitudinal electric field of order $E_{pw}$ (which can also be
written as $m_e c^2 / e \lambdabarp$.)
The transverse momentum of the jet particles broadened significantly
from the initial distribution.
Some of the initial jet energy was also converted into heating the
background plasma, with the thermal energy of plasma electrons 
increased to approximately $m_e c^2$.

To explore the generality of our results, we varied the physical 
parameters in our
simulations as described earlier.  We observe inductive and wakefield 
acceleration
for the parameter regime where the Weibel instability occurs.
Both continuous and finite-length jets exhibit the same transverse (Weibel) 
dynamics but different longitudinal dynamics.
The longitudinal electrostatic fields are  much stronger in the
finite-length case.  
Varying the jet diameter does not affect the acceleration mechanisms that we
observe.  As the jet diameter is increased, we observe the
same expulsion of positron filaments from the surface as in the narrower jet
case, as well as the interior filaments of the same sign coalescing within
the jet.

For a finite length jet, the maximum $E_z$ is approximately 
proportional to $\alpha$ for $\alpha < 1$.  This linear behavior has been
observed in previous studies of plasma wakefield 
accelerators~\cite{Esarey96}.  
For $\alpha >1$, the scaling we observe is approximately $\alpha^{0.5}$.
We also observe almost complete expulsion 
of the background plasma electrons by the jet and formation of a 
focusing ion channel.
This case is similar to the so-called
``blow-out'' regime previously studied in electron beam-plasma 
interactions~\cite{Esarey96,Lee2000}.

The density ratio and jet temperature are expected to affect the dynamics 
of the Weibel instability~\cite{Silva2002}.  
For a given density ratio, increased temperature
provides a pressure that opposes the transverse magnetic pinching force.
The threshold condition for instability is approximately 
$\alpha > \gamma (\Dbetaperp)^2$ for a relativistic jet.  
This threshold condition was satisfied over nearly the entire range of our 
simulation parameters. The range of $\Delta \beta_{\perp}$ in our study is
similar to those of earlier studies of Weibel instability in 
astrophysical plasmas
~\cite{silva03,frederiksen04,hededal04,nishikawa05}.
To test the instability criteria, we simulated a case with
$\alpha = 0.1$ and $\Dbetaperp$ approximately equal to 0.1 for $\gamma = 10$.  
No Weibel instability was observed over a period of $180~\invomegape$.
The implication is that for hot, tenuous jets, Weibel filamentation
is suppressed.

The Weibel instability is believed to occur in astrophysical environments 
where relativistic outflows exist~\cite{Medvedev99,Medvedev05,
silva03,frederiksen04,hededal04,nishikawa05}.
Our results indicate that when the Weibel instability occurs, 
an inductive ``Faraday acceleration'' 
mechanism can effectively power cosmic particles, and electrostatic 
plasma wakefields are excited by an
initially charge-neutral jet when oppositely charged 
filaments separate.  We expect the longitudinal electric fields
observed in our simulation will continue to be generated as the jet
propagates over long distances.
Relativistic particles will be accelerated as long as they
remain in phase with these waves.
Due to the velocity difference, acceleration 
occurs for at most one half period of phase slip between the particle at
velocity $c$ and the slower wave at $0.995c$ created by the $\gamma = 10$
jet.  The dephasing length is approximately 
$\gamma^2 \lambda_p = 200 \pi \lambdabarp$, and
thus the energy gain is about 300~MeV
for an accelerating field of order $E_{pw}$.
This is a plausible mechanism for creating a
population of relativistic particles that give rise to the observed 
gamma-rays.  It could also serve as the first stage injection mechanism
for other ultra-high energy cosmic ray acceleration models.

These simulations have implications for future research.  The jet-plasma 
interaction should be followed for a much longer
time to study the effect of particle acceleration over long distances.
The effect of background magnetic fields should be investigated, 
as well as different particle mixtures of
jet and background plasma compositions.  Synchrotron radiation can be readily
implemented in a particle-in-cell code and the resulting spectrum can be
verified against observation.
Advances in accelerator technology will enable
jet-plasma dynamics to be studied in a terrestrial laboratory 
environment~\cite{ng03}.
Experimental results will provide important tests of simulations,
thus improving the connection between simulations and observations.
The work presented here provides the basis for these experiments.

We appreciate discussions with K.-I.~Nishikawa, K.~Reil, A.~Spitkovsky, 
and M.~Watson. We would also like to thank P.~Chen, R.~Ruth,
and R.~Siemann for their support and encouragement.
Work supported by the U.S. Department of Energy
under contract number DE-AC02-76SF00515.

\bibliography{ng_noble}

\end{document}